\documentclass[12pt]{article}
\usepackage{a4wide}
\usepackage{graphicx}
\usepackage{amsfonts}

\def\a{\alpha}
\def\r{\rho}
\def\s{\sigma}
\def\p{\partial}
	
\def\bm#1{\mbox{\boldmath{$#1$}}}

\def\H{{\cal L}}

\def\ra{\rightarrow}

\begin{document}

\title{Renormalization Group and Quantum Information}

\author{Jos\'e Gaite\\
{\small\em Instituto de Matem{\'a}ticas y F{\'\i}sica
Fundamental,}
{\small\em CSIC, Serrano 113 bis, 28006 Madrid, Spain}
}
\date{March 22, 2006}  
\maketitle

\begin{abstract}
The renormalization group is a tool that allows one to obtain a
reduced description of systems with many degrees of freedom while
preserving the relevant features.  In the case of quantum systems, in
particular, one-dimensional systems defined on a chain, an optimal
formulation is given by White's ``density matrix renormalization
group". This formulation can be shown to rely on concepts of the
developing theory of quantum information. Furthermore, White's
algorithm can be connected with a peculiar type of quantization,
namely, {\em angular quantization}. This type of quantization arose in
connection with quantum gravity problems, in particular, the {\em
Unruh effect} in the problem of black-hole entropy and Hawking
radiation.  This connection highlights the importance of quantum
system boundaries, regarding the concentration of quantum states on
them, and helps us to understand the optimal nature of White's
algorithm.
\end{abstract}
     
\section{Introduction}

The renormalization group arose in quantum field theory as a
transformation of the coupling constant(s) equivalent to a resummation
of perturbation theory. It was later generalized by Wilson to
statistical systems as a transformation of the full probability
distribution, which is defined by an unbounded set of parameters.  One
of the most interesting aspects of this general renormalization group
is that it can be understood as a transformation that removes
small-scale degrees of freedom but preserves thet set of degrees of
freedom relevant to describe overall features, such as they are needed
in the description of phase transitions, for example.  In that sense,
a renormalization group transformation is not reversible, so we should
speak of a {\em semigroup} rather than of a group.

Regarding quantum many-body systems, one is most interested in
calculating relevant features of the ground state (and maybe some
excited states). This is a problem suitable for a renormalization
group treatment, and indeed related to the problem of calculating
the properties of classical statistical systems.  There are many
particular renormalization group algorithms that remove small-scale
degrees of freedom in various ways. In principle, any algorithm that
approaches a fixed point is valid. However, the rate of convergence
can vary broadly. It has been a problem to find efficient
algorithms. In this regard, a deeper understanding of the process of
removal of small scale information surely helps. Here we shall focus
on White's ``density matrix renormalization group"
\cite{White,Scholl}, which has proved to be very powerful and which is
indeed based on a deep analysis of the the process of removal of small
scale information.

The analysis of the process of removal of small scale information is
actually part of the analysis of information processing and,
therefore, belongs to information theory.  Furthermore, an analogy can
be established between renormalization group transformations and the
{\em irreversible evolution} of statistical systems, as implied by
Boltzmann's $H$-theorem \cite{I-OC}.  Since the density matrix
renormalization group is an optimal tool for quantum systems (albeit
mainly one-dimensional chains), we can expect a strong link with the
theory of quantum information.  This theory has its origins in work
done early in the past century, shortly after the discovery of quantum
theory itself. However, it has only undergone a period of rapid
development during the last years, in relation with the prospects of
quantum computation.  The semigroup character of the quantum
renormalization group is best understood by using concepts of quantum
information theory, as we shall do.

White's algorithm involves a doubling of the system at each
renormalization group iteration. As we will see, this makes sense from
the quantum information standpoint and, in addition, has an
interesting interpretation, since there is a relation with angular
quantization \cite{I}. Indeed, White's algorithm can be connected with
another way of solving quantum systems; namely, quantum chains can be
solved by relating them with classical two-dimensional systems on a
lattice and using the {\em corner transfer matrix} method.  The
continuum limit gives rise to a peculiar type of quantization, namely,
{\em angular quantization}, valid for {\em relativistic} quantum field
theories, but different from the standard canonical quantization. This
type of quantization was introduced in connection with quantum gravity
problems, namely, the problems of black-hole entropy and Hawking
radiation \cite{BD}.  Angular quantization yields the relevant states
in the calculation of the density matrix, showing precisely how the
full spectrum is truncated to remove small-scale degrees of freedom
\cite{I}. In particular, it renders transparent the importance of
quantum system boundaries, where quantum states concentrate.

Some new developments in the area of  
renormalization group and quantum information have appeared 
recently \cite{new}.

\section{The density matrix renormalization group}

Strongly correlated electron systems have become an important subject
in condensed matter physics. This has led to the development of
suitable approximation methods for quantum lattice models. Chiefly
among them is the renormalization group, which has the philosophy of
truncating the multitude of states to the {\em relevant} ones to
describe the physical properties of the system in certain domain.
There are many ways to implement this idea (not all of which can be
properly called renormalization group methods).  The various
formulations of the quantum renormalization group have different
efficacy, depending on the particular model to which they are
applied. Two succesful classical approaches are Wilson's treatment of
the Kondo impurity problem and Kadanoff's blocking technique.  These
methods belong to the the class of numerical renormalization groups in
{\em real space} (as opposed to Fourier space).  Wilson's treatment of
the Kondo problem is accurate, but it relies on the special nature of
the interaction in it. The blocking technique is universal for lattice
models but rather inaccurate, as we analyse next.

\subsection{Problems with the real space renormalization group}

\begin{figure}
%\centering{\includegraphics[width=8cm]{../RG/real-sp.eps}}
\centering{\includegraphics[width=8cm]{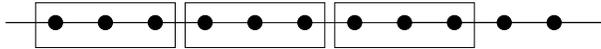}}
\caption{Blocking of a one-dimensional chain, with three sites per block.}
\label{blocks}
\end{figure}

Consider a one-dimensional lattice model with $N$ sites. If $N$ is
sufficiently large to represent a realistic system, the Hamiltonian
involves a huge number of states, because the total number of states
grows exponentially with $N$. There is no possibility to diagonalize
such a large Hamiltonian. Instead, we may break the chain into a
number of equal size blocks, and we treat a block as a small size
system (see Fig.\ \ref{blocks}). We can then diagonalize the block
Hamiltonian and discard the higher energy states. Each block, with the
states kept on it, can now be considered as a site of new system with
a fraction of the initial $N$ sites. In so doing, we get an effective
or {\em renormalised} interaction between sites. This procedure can be
iterated until the the initial size $N$ is reduced to a small number.

However, the blocking RG converges slowly. White and Noack \cite{W-N}
realized that the problem lies in the choice of block eigenstates as
the states to be kept: these states belong to a small system, namely,
the block, in which the boundary conditions are very important. In
other words, isolating a block from its neighbours destroys the
quantum correlations between them, which are very important for the
low-energy spectrum of the total system (with $N$ sites). These
correlations are somehow recovered by the renormalization of the
couplings, but in a small amount. A partial remedy is the
``combination of boundary conditions" approach \cite{W-N}, namely, to
consider block states corresponding to various boundary
conditions. This approach is effective in some cases only.

So White and Noack \cite{W-N} proposed to diagonalize a larger block,
the ``superblock", which includes the basic block. Then the problem is
to project the ``superblock" state (or states) onto block states: we
need a criterium to select which ones to keep. White's intuition led
him to appeal to Feynman's philosophy on the density matrix formalism:
a density matrix simply represents the correlation of a quantum system
with the rest of the universe. This correlation is usually called {\em
entanglement}. The conclusion White drew is that the block states to
be neglected, among the density matrix eigenstates, are the ones with
small eigenvalues, because they hardly contribute to physical
observables \cite{White}.  Let us recall White's procedure
\cite{White} more precisely.

\subsection{Density matrix renormalization group algorithm}    
\label{algo}

Let us have a one-dimensional quantum system on a chain (finite or
infinite).  We select a relatively large block (the ``superblock") but
such that it can be exactly diagonalized. We obtain its ground state
$|\psi\rangle$, which we take as {\em environment} of a smaller block
included in the superblock. Let $|i\rangle,\;i=1,\ldots,\ell,$ be a
complete set of block states and $|j\rangle,\;j=1,\ldots,J,$ be the
states of the rest of the ``superblock" (Feynman's rest of the
universe). Then $|\psi\rangle =
\sum_{i,j}\psi_{ij}|i\rangle|j\rangle.$ We want to find a subset of
block states $|a\rangle,\;a=1,\ldots,m <\ell,$ such that they provide
an optimal reduced representation of the block in the environment
(boundary conditions) given by the superblock state.  In other words,
we want $|\tilde\psi\rangle =
\sum_{a,j}\tilde\psi_{aj}|a\rangle|j\rangle$ to be as close to
$|\psi\rangle$ as possible. White's prescription is to minimize the
``distance"
\begin{equation}
S = | |\psi\rangle - |{\tilde\psi}\rangle |^2.
\label{WhiteS}
\end{equation}
Since both $|\psi\rangle$ and $|{\tilde\psi}\rangle$ are actually
matrices, this distace is in fact the standard distance between
matrices.  He shows that this minimization problem amounts to the
{\em singular value decomposition} of the rectangular matrix
$\psi_{ij}$. One writes $\psi = UDV^T$, where $U$ and $D$ are
$\ell\times \ell$ matrices, $V$ is an $\ell\times J$ matrix ($J \geq
\ell$), $U$ is orthogonal and $V$ column orthogonal, and $D$ is a
diagonal matrix containing the singular values. The arbitrary integer
$m <\ell$ defines the number $\ell - m$ of singular values to be
neglected. Actually, $U$ is the matrix formed by the eigenvectors of
the block density matrix and $\r = U D^2 U^T.$ Removing $\ell - m$
singular values si equivalent to keeping the $m$ most important
eigenvectors $|a\rangle$ of the block density matrix.

The construction of an iterative algorithm that implements the
previous result is not difficult. A convenient algorithm \cite{White},
inspired by Wilson's treatment of the Kondo problem, can be
schematically expressed as follows:
  \begin{enumerate}
\item Select a sufficiently small, soluble block $[0,L]$:
%\hspace{1cm}\includegraphics[width=4cm]%
%{D:/export/home/gaitecj/semin/TH02/chain.eps}
\hspace{1cm}\includegraphics[width=4cm]{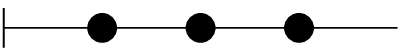}
\item Reflect the block on the origin: 
%\hspace{1cm}\includegraphics[width=7cm]%
%{D:/export/home/gaitecj/semin/TH02/chain2.eps}
\hspace{1cm}\includegraphics[width=7cm]{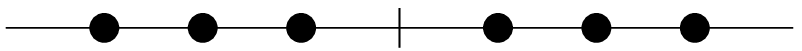}
\item Compute the ground state.
\item Compute the density matrix of the block $[0,L]$.
\item Discard eigenstates with smallest eigenvalues.
\item Add one site next to the origin.
\item Go to 2.
  \end{enumerate}
(This algorithm is to be iterated indefinitely and represents the
``infinite-system algorithm", but there is also a ``finite-system algorithm". 
We refer the reader to White's papers \cite{White} for more details.)
One has to adjust this procedure in such a way that the iteration
keeps the Hilbert space size approximately constant. The procedure can
be performed algebraically for a chain of coupled {\em harmonic}
oscillators \cite{I}. Otherwise, it has to be performed numerically.

\subsubsection{Density matrix renormalization group for mixed states}    

We have assumed so far that the system is in a pure state (the ground
state), which we want to calculate. One can also consider mixed states
\cite{White}.  In particular, it is useful to consider the properties
of thermal states. If we represent the mixed state by means of a set
of Boltzmann weights $w_k$, then we have to minimize
\begin{equation}
S = \sum_k w_k | |\psi^k\rangle - |{\tilde\psi^k}\rangle |^2,
\label{WhiteSm}
\end{equation}
where 
${\tilde\psi^k}_{ij} = \sum_{\a,j}a_{\a}^k u^{\a}_i v_j^{k,\a}.$
The optimal solution is again to neglect the smallest (most singular)
eigenvalues of the density matrix
$$\r_{ii'} = \sum_k w_k \sum_j \psi^k_{ij}\,\psi^k_{i'j}\,,$$
namely, the smallest values of $\sum_k w_k |a_{\a}^k|^2.$
The same type of iterative algorithm can still be used.

\section{Entanglement entropy and quantum information}

{\em Entanglement} or non-separability refers to the existence of
quantum correlations between two sets of degrees of freedom of a
physical system that can be considered as subsystems
\cite{Peres,Preskill}.  Two (sub)systems in interaction are entangled
and their entanglement continues after their interaction has
ceased. This fact gives rise to the Einstein-Podolsky-Rosen (EPR)
paradox.  So, while there can be entanglement without interaction,
interaction always produces entanglement.

It is clear that entanglement plays a r\^ole in the density matrix
renormalization group and, in general, in quantum phase transitions.
This has been realized recently, by researchers in quantum information
theory \cite{Os-Ni,Vi-Lato}. The subject linking quantum information
theory to traditional problems in condensed matter theory surely
deserves further study.  Here we review the concept of entanglement
and other relevant concepts of information theory, regarding their
r\^ole in the density matrix renormalization group.

\subsection{Entanglement entropy of a bipartite system}

The entanglement of two parts of a quantum system can be measured by
the von Neumann entropy. This entropy is defined in terms of the
density matrices of each part. We may consider, for later convenience,
one part as ``left'' and another as ``right'' or one part as
``exterior'' and another as ``interior''.  Then, let us represent
states belonging to the left part with small letters and states
belonging to the right part with capital letters. A basis for the
global states (left plus right) is
$\{|a\rangle\}\otimes\{|A\rangle\}$. Let us take a global state, say
the ground state for definiteness, and represent it in this basis as
\begin{equation}
|0\rangle=\sum_{aA} \psi_{aA}\, |a\rangle \otimes |A\rangle,
\end{equation}
defining a coefficient matrix $\psi_{aA}$.
Then we have two {\em different} density matrices, for each part:
\begin{equation}
\rho_{L}=
\frac{ {\psi}^*{\psi}^{{^T}}}{ {\rm Tr}\,{\psi}^*{\psi}^{{^T}} }\,,
\quad
\rho_{R}=
\frac{ {\psi}^\dagger{\psi}} {{\rm Tr}\,{\psi}^\dagger{\psi} }\,. 
\end{equation}
Correspondingly, we have two von Neumann entropies:
\begin{equation}
S_{L}=-{\rm Tr}_a\left(\rho_{L}\ln \rho_{L}\right),\quad
S_{R}=-{\rm Tr}_A\left(\rho_{R}\ln \rho_{R}\right).
\end{equation}
Now it is important to recall the ``symmetry theorem'', which states
that both entropies are equal, $S_{L}= S_{R}$. This can be proved in
several ways; for example, by using the {\em Schmidt decomposition} of
the entangled state: both $\rho_{L}$ and $\rho_{R}$ have the same
non-zero eigenvalues \cite{Preskill}. Let us remark that the Schmidt
decomposition embodies entanglement and, actually, the singular value
decomposition (as used by White) is its finite-dimensional version.
The equality of entropies may seem somewhat paradoxical, since there
can be many more degrees of freedom in one part (the exterior or rest
of the universe) than in the other.  But the entanglement entropies
are associated with properties shared by both parts, that is, with
(quantum) correlations.

Let us see how interaction produces entanglement and increases the
entropy.  Consider two non-interacting parts of a quantum system that
are originally in respective mixed states. After their interaction,
which we describe as an arbitrary unitary evolution of the composite
system, the initial density matrix $\rho_{L} \otimes \rho_{R}$ has
evolved to $\rho'_{LR}$. As a consequence of {\em subadditivity} of
the entropy, it is easy to see that the partial traces $\rho'_{L}$ and
$\rho'_{R}$ have in general von Neumann entropies $S'_{L}$ and
$S'_{R}$ such that $S'_{L} + S'_{R} \geq S_{L} + S_{R}$
\cite{Peres,Preskill}. Of course, if the initial state is a product of
pure states, $S_{L} = S_{R} = 0.$ In essence, this increase of entropy
after interaction is an abstract form of the second law of
thermodynamics.

\subsection{Information theory and maximum entropy principle}

The entropy concept arose in thermodynamics but only took a truly
fundamental meaning with the advent of information theory. In this
theory, entropy is just uncertainty or missing information, while
information itself is often called {\em negentropy}. We recall basic
definitions: the information attached to an event that occurs with
probability $p_n$ is $I_n = -\log_2 p_n$ (measured in {\em bits});
therefore, the average information (per event) of a source of events
is $$S(\{ p_n \}) = \sum_n p_n I_n = - \sum_n p_n \log_2 p_n\,.$$ This
average information is called the entropy of the source.  Note that
improbable events convey more information but contribute less to the
entropy: $p\,I(p) = -p\,\log_2 p$ is concave and has its maximum at
$p=e^{-1}$.

The previous definitions, given by Shannon in his theory of
communication, may seem unrelated to thermodynamic entropy as a
property of a physical system. However, according to the foundations
of Statistical Mechanics on Probability Theory (the Gibbs concept of
ensembles), a clear relation can be established. This was done by
Jaynes \cite{Jay}, by appealing to the Bayesian philosophy of
probability theory. In this philosophy, the concept of ``a priori''
knowledge is crucial.  Indeed, although the exact microscopic state of
a system with many degrees of freedom may be unknown, one has some ``a
priori'' knowledge given by the known macroscopic variables.  Jaynes
postulates, according to Bayesian philosophy, that the best
probability distribution to be attributed to a stochastic event is
such that it incorporates only the ``a priori'' knowledge about the
event and nothing else.  This postulate amounts to Jaynes' {\em
maximum entropy} principle: given some constraints, one must find the
maximum entropy probability distribution (density matrix, in the
quantum case) compatible with those constraints, usually, by
implementing them via Lagrange multipliers.  In particular, more
constraints mean less missing information and so less entropy.

\subsection{Information geometry}

{\em Distinguishability} of probability distributions is an important
concept in information theory.  The question is when two probability
distributions are sufficiently distinguishable for some purpose and
what measures are necessary to distinguish them. This in an important
problem in statistics, in particular, in estimation theory. It has led
to endow spaces of probability distributions with a metric geometry.
 
Let us briefly review the fundamental concepts of information geometry
\cite{Amari}. Let $p(x,\xi)$ be an $n$-parameter family of probability
distributions ($\xi \in \mathbb{R}^n$).  The primordial concept is the
existence of a metric, namely, the {\em Fisher information matrix}:
\begin{equation}
g_{ij}(\xi) = 4 \int \p_i \sqrt{p(x,\xi)}\, \p_j \sqrt{p(x,\xi)}\, dx
\label{Fisher}
\end{equation}
(the derivatives are taken with respect to the parameters).  This
metric provides any space of probability distributions with a
Riemannian structure. Hence, one can introduce the $\a$-{\em
connections}, the case $\a = 0$ being the standard Riemannian
connection with respect to the Fisher metric.  The next important
concept is the notion of {\em divergence function}. It is a real
positive function of a pair of probability distributions that vanishes
if both distributions in the pair coincide. So divergences are
distance-like measures, but they do not satisfy in general the
remaining axioms of distance, in particular, they are not in general
symmetric. However, a divergence's differential form is in fact
symmetric and constitutes a metric.  Most important are the
$\a$-divergences, given (in the discrete case) by
\begin{eqnarray}
D^{(\a)}(p_i,q_j) =  \sum_i p_i f(q_i/p_i),\\
f(x) = \left\{
\begin{array}{cl}
\frac{4}{1-\a^2} \left(1 - x^{(1+\a)/2}\right)  
& \a \neq \pm 1 \\ - \ln x & \a = -1 \\ x \ln x & \a = 1\,.
\end{array}
\right.
\end{eqnarray}
They are related to generalized entropies: R\'enyi's $\a$-entropies,
Tsallis' entropy, etc.

In general, opposite sign divergences satisfy 
\begin{equation}
D^{(-\a)}(p_i,q_j) = D^{(\a)}(q_i,p_j),
%\label{}
\end{equation}
and are called {\em dual}.  In particular, the case $\a = 0$ is
symmetric and actually provides a distance, namely,
$\sqrt{D^{(0)}(p_i,q_j)}$, with
\begin{equation}
D^{(0)}(p_i,q_j) = 2 \sum_i \left(\sqrt{p_i} - \sqrt{q_i}\right)^2.
\label{Hell}
\end{equation}
The $\pm 1$-divergence is called the Kullback-Leibler divergence or
{\em relative entropy} and is particularly important. Its differential
form yields the Fisher metric in the continuous case.

Note that the distance defined by Eq.\ (\ref{Hell}) has a simple 
interpretation. To see it, let us associate with a probability 
distribution the vector $\{\sqrt{p_i}\}$, so that the probability 
normalization becomes a vector normalization. Therefore, 
probability distributions become rays in a real vector space 
\cite{Bro-Hug}. The distance $\sqrt{D^{(0)}(p_i,q_j)}$ is just
the standard distance in this vector space, or rather the induced 
distance in the corresponding projective space. In the continuous case,
the standard Euclidean metric induces a metric in the parameter 
manifold that is precisely the Fisher metric (\ref{Fisher}).

The above defined concepts are classical but they all admit quantum
generalizations \cite{Amari}.  The square root of the $\a=0$ quantum
divergence coincides with the {\em Bures distance} between density
matrices.  This distance, restricted to the subspace of diagonal
density matrices (for a fixed basis), does indeed become simply the
square root of $2\,\mathrm{Tr} (\r_1^{1/2} - \r_2^{1/2})^{2}$
\cite{Hubner}.  In general, we can write $\r_1 = W_1\,W_1^\dagger,$
$\r_2 = W_2\,W_2^\dagger$ for different pairs of operators
$\{W_1,W_2\}$, and the Bures distance can be defined by the infimum
\begin{equation}
D^{(0)}(\r_1,\r_2) = 2\,\mathrm{inf~Tr} (W_1 - W_2)(W_1 - W_2)^\dagger.
\label{Bures}
\end{equation}
For pure states, the Bures distance is just the natural distance in
the complex projective Hilbert space, such that its infinitesimal form
is the Fubini-Study metric \cite{Woo,Gib,BraCa}.  Notably, the
distance between mixed states (density matrices) is given by
minimizing the distance between their respective {\em purifications}
in a larger Hilbert space \cite{Jo}.

\subsection{Quantum information theory}

The concepts of Shannon's classical theory of communication have
quantum analogues \cite{Schu,Preskill}. But the quantum theory of
communication is richer. Indeed, the key new notion in the quantum
theory is entanglement (as already described).  Schumacher studied the
problem of quantum coding and, in particular, the problem of
communication of an entangled state \cite{Schu}. The technical name is
{\em transposition}, since the copy of a quantum state is not
possible, a fact that constitutes the no-cloning theorem
\cite{Peres}. His conclusion was that the von Neumann entropy of the
state is the quantity that determines the {\em fidelity} of the
transposition: it is possible to transpose the state with near-perfect
fidelity if the signal can carry at least that information.  The
method is analogous to classical coding, that is, one is to discard
small probabilities, but involves the use of the Schmidt
decomposition.

Of course, fidelity and distinguishability are related concepts, and
indeed the fidelity $F(\r_1,\r_2) = (1 - D^{(0)}(\r_1,\r_2)/4)^2$
\cite{Jo}.  Maximal fidelity ($F = 1$) is equivalent to perfect
indistinguishability.  Minimal fidelity ($F = 0$) takes place between
maximally separated mixed states, since the Bures distance is bounded
(this is obvious for pure states).

\subsection{Quantum information interpretation of the 
density matrix renormalization group}

Schumacher's approximate transposition of an entangled state is
essentially identical to White's procedure.  Moreover, White's
distance criteria can be interpreted in terms of the distances defined
in information geometry.

Let us recall White's prescription: select the block states to be kept
by minimizing a ``distance" $S = | |\psi\rangle - |{\tilde\psi}\rangle
|^2$ between the actual superblock state $|\psi\rangle$ and its
approximation $|{\tilde\psi}\rangle$ [Eq.\ (\ref{WhiteS})].  In terms
of quantum information theory, we want to maximize the fidelity of the
block mixed state or, in other words, to minimize the distance between
the actual and approximated block mixed states. According to Jozsa's
result \cite{Jo}, this minimization can be achieved by minimizing the
distance between their respective {\em purifications} in a larger
Hilbert space, which in this case is just the superblock.  So it is
correct to minimize de Hilbert space distance $S$.

The density matrix renormalization group for mixed states also
corresponds to a distance minimization, in the space of superblock
mixed states. The reference state is $\sum_k w_k |\psi^k\rangle
\langle\psi^k|$ and its approximation $\sum_k w_k |\tilde\psi^k\rangle
\langle\tilde\psi^k|$. Then, according to Eq.\ (\ref{Bures}),
$$D^{(0)}(\r_1,\r_2) = 2\sum_k w_k | |\psi^k\rangle -
|{\tilde\psi^k}\rangle |^2.$$

\section{White's algorithm and angular quantization}

White's density matrix renormalization group algorithm, exposed in
Sect.\ \ref{algo}, can be purely justified on a quantum information
basis as follows.  If $\r_1$ and $\r_2$ are two mixed states of a
Hilbert space ${\cal H}$, then ${\cal H} \otimes {\cal H}$ is the
smallest Hilbert space that contains purifications of both states
\cite{Jo}. Therefore, for a block of given size, the most economical
``rest of the universe'' is a reflection of the block (with the same
size).

However, the particular geometry in White's algorithm lends itself to
a more fruitful connection, namely, the connection with angular
quantization \cite{I}. Before explaining angular quantization, we must
introduce the corner transfer matrix, a method of solving
two-dimensional classical systems that turns out to be related to
White's algorithm.

\subsection{Corner transfer matrix and density matrix}
\label{CTM}

Let us first recall the connection between quantum 
mechanics and classical statistical mechanics in one more
dimension, realized by the Euclidean path integral.
For spin chains, the equivalent classical system is
defined on a two-dimensional lattice and the partition function can be
conveniently expressed in terms of the {\em transfer matrix}. This
matrix evolves the system from one row to the next one. In addition to
the rwo-to-row transfer matrix, there was defined the {\em corner
transfer matrix} (in the context of soluble models).  This matrix
evolves the system from one side of the corner to the other side.  The
formulation by Baxter of the corner transfer matrix for soluble models
is old, but its importance in our context was realized later, in a
paper by Thacker \cite{Thac}, in which he showed that the relevant
symmetry is best understood in the continuum limit, as we shall see.

\begin{figure}
\centering{%
\includegraphics[width=6cm]{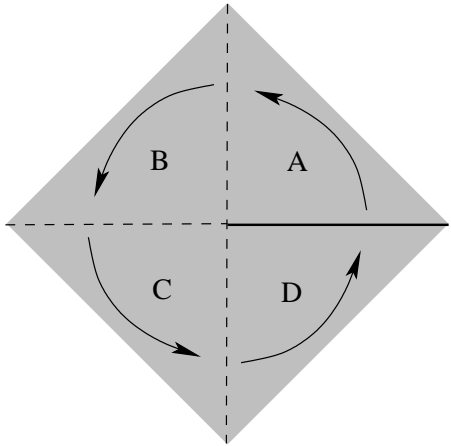}%
}
\caption{Composition of four corner transfer matrices in the direction
of the arrows.}
\label{ABCD}
\end{figure}

To introduce the corner transfer matrix, a site in the middle of the
two-dimensional lattice is chosen as the origin, and then the spins
(or other site variables) are fixed along the vertical and horizontal
axes.  Four different corner transfer matrices, say $A,B,C,D,$ are
defined by summing over the remaining site variables in each quadrant.
Then the partition function is $Z = \mathrm{Tr} (ABCD).$ The matrix
$ABCD$ represents the transfer from one side of the right horizontal
semiaxis to the other side, as shown in Fig.\ \ref{ABCD}.  If we
define the state on the horizontal axis by the vector
$\Psi({\bm\s}_L,{\bm\s}_R)$ (splitting it on both semiaxes), we have
$$ABCD = \sum_{\s_L}
\Psi^*({\bm\s}_L,{\bm\s}_R)\Psi({\bm\s}_L,{\bm\s}'_R)\,.$$ Of course,
this matrix (with entries ${\bm\s}_R$ and ${\bm\s}_r'$) corresponds to
the density matrix of the right horizontal semiaxis in the evironment
provided by the the left horizontal semiaxis. This connection was
realized by Nishino and Okunishi \cite{Nishi} and developed by Peschel
and collaborators \cite{Peschel}.

On an isotropic lattice, the four corner transfer matrices can be
arranged to coincide, so
$$\r_R = A^4 = \exp\left(-H_\mathrm{CTM}\right),$$ defining a sort of
corner-transfer-matrix Hamiltonian, such that $Z =
\mathrm{Tr}\,\exp\left(-H_\mathrm{CTM}\right)$ \cite{Peschel}. Roughly
speaking, this Hamiltonian adopts the form $H_\mathrm{CTM} = \sum_n n
\,H_n$, where $H_n$ is a quasi-local Hamiltonian and the index $n$
runs over sites. In comparison with the standard Hamiltonian for the
rwo-to-row transfer matrix, we remark that the low ``energy''
contribution of local states is depressed as we move away form the
origin, due to the factor $n$ (and viceversa).  To substantiate this
intuitive picture, we need to explain angular quantization in the
continuum limit (field theory).

\subsection{Field theory half-space density matrix}

Let us consider, for definiteness, a chain of coupled oscillators.  In
the continuum limit, that is, for large correlation length, the model
becomes simpler, in spite of having a non-denumerable set of degrees
of freedom.  The action for this model, namely, a one-dimensional
scalar field, is (after a few redefinitions)
\begin{equation}
A[\varphi(x,t)] = \int\! dt\, dx \left({1\over 2}
\left[(\partial_t\varphi)^2 -
(\partial_x \varphi)^2 \right] - V(\varphi) \right),
\label{I}
\end{equation}
where $\varphi$ is the field. So the chain is described by a {\em
relativistic} 1+1 field theory (relativistic with respect to the sound
speed, normalized to one).

Let us obtain a path integral representation for the density matrix on
the half-line with respect to the ground state (the vacuum) of action
(\ref{I}) \cite{Bomb,KabStr,CalWil}.  In the continuum limit, the
half-line density matrix is a functional integral,
\begin{equation}
\r[\varphi_R(x),\varphi'_R(x)] =
\int\! D\varphi_L(x)\, \psi_0[\varphi_L(x),\varphi_R(x)]\,
\psi^*_0[\varphi_L(x),\varphi'_R(x)],
\label{DM0}
\end{equation}
where the subscripts refer to the left or right position of the
coordinates with respect to the boundary (the origin). Now, we must
express the ground-state wave-function as a path integral,
\begin{equation}
\psi_0[\varphi_L(x),\varphi_R(x)] = \int D\varphi(x,t)\,
\exp\left(-A[\varphi(x,t)]\right),
\end{equation}
where $t\in (-\infty,0]$ and with boundary conditions $\varphi(x,0) =
\varphi_L(x)$ if $x<0$, and $\varphi(x,0) = \varphi_R(x)$ if $x>0$.
The conjugate wave function is given by the same path integral and
boundary conditions but with $t\in [0,\infty)$.  Substituting into
Eq.~(\ref{DM0}) and performing the integral over $\varphi_L(x)$, one
can express $\r(\varphi_R,\varphi'_R)$ as a path integral over
$\varphi(x,t)$, with $t\in (-\infty,\infty)$, and boundary conditions
$\varphi_R(x,0+) = \varphi'_R(x)$, $\varphi_R(x,0-) = \varphi_R(x)$.
In other words, $\r(\varphi_R,\varphi'_R)$ is represented by a single
path integral covering the entire plane with a cut along the positive
semiaxis, where the boundary conditions are imposed.

Next, we need to calculate the density matrix, which we can do by
diagonalizing it in the appropriate basis.

\subsection{Angular quantization and Rindler space}

Two-dimensional relativistic field theory has Lorentz symmetry, 
which becomes just rotational symmetry in its Euclidean version.
The generator of rotations in the $(x,t)$ plane is given by
\begin{equation}
\H =  \int dx\,(x\,T_{00}-t\,T_{11}),
\label{L}
\end{equation}
in terms of the components of the stress tensor computed from the
action (\ref{I}). Of course, $T_{00}$ is the Hamiltonian density and
$T_{11}$ the momentum density.  To simplify, one can evaluate $\H$ at
$t=0$, obtaining a Hamiltonian that we recognize as the continuum
limit of $H_\mathrm{CTM}$, defined in Sect.\ \ref{CTM}.

For quantization, let us consider a free action [$V(\varphi)=0$].  In
the Schr\"odinger representation, we should replace the momentum $\Pi
= \partial_t\varphi$ with $\Pi(x) = i\,\delta/\delta \varphi(x)$.
However, as in canonical quantization, one rather uses the
second-quantization method, which diagonalizes the Hamiltonian by
solving the classical equations of motion and quantizing the
corresponding normal modes.  Let us recall that, in canonical
quantization, if we disregard anharmonic terms, the classical
equations of motion in the continuum limit become the Klein-Gordon
field equation, giving rise to the usual Fock space.  
In an angular analogy, the eigenvalue equation for $\H$ leads to the
Klein-Gordon equation in polar coordinates in the $(x,t)$ plane,
The free field wave equation in polar coordinates,
\begin{equation}
(\Delta + m^2) \varphi = \left({1\over r}{\partial \over \partial
r}r{\partial \over \partial r} + {1\over r^2}{\partial^2 \over
\partial \phi^2}+ m^2\right) \varphi = 0,
\end{equation}
can be solved by separating the angular variable: it becomes a Bessel
differential equation in the $r$ coordinate with complex solutions
$I_{\pm i\,\ell}(m\,r)$, $\ell$ being the angular
frequency. We have a continuous spectrum, which becomes discrete on
introducing boundary conditions. One of them must be set at a short
distance from the origin, to act as an ultraviolet regulator
\cite{Bomb,KabStr,CalWil}, necessary in the continuum limit.

Therefore, the second-quantized field is (on the positive semiaxis
$t=0 \Leftrightarrow \phi = 0$, $x \equiv r$)
\begin{equation}
\varphi(x) = \int {d\ell\over 2\pi}\,\frac{b_\ell\,I_{i\,\ell}(m\,x) +
b_\ell^\dag\,I_{-i\,\ell}(m\,x)}{\sqrt{2\,\sinh(\pi\,\ell)}},
\end{equation}
where we have introduced annihilation and creation operators and where
the term that appears in the denominator is just for normalization, to
ensure that those operators satisfy canonical conmutations relations.
There is an associated Fock space built by acting with $b_\ell^\dag$
on the ``vacuum state''. These states constitute the spectrum of
eigenstates of $\H$, which adopts the form $\H = \int d\ell \,\ell\,
b_\ell^\dag b_\ell$ (where the integral is replaced with a sum for
discrete $\ell$). They are the density matrix eigenstates as well.

\begin{figure}
\centering{%
\includegraphics[width=9cm]{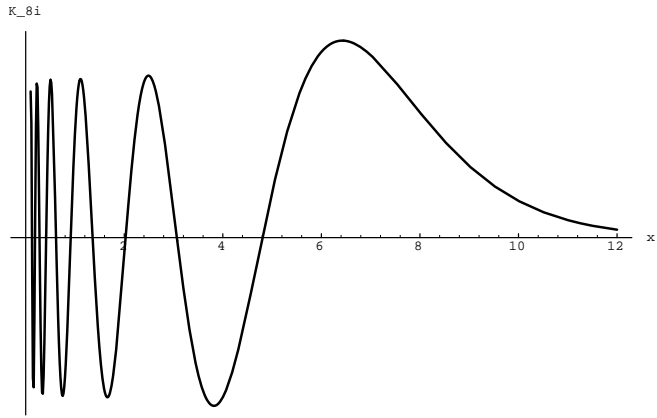}%
}
\caption{Angular-quantization wave $K_{8\,i}(x)$. Note the behaviour near 
$x=0$.}
\label{wave}
\end{figure}

Let us remark that the functions $I_{\pm i\,\ell}(m\,x)$ have
wave-lengths that increase with $x$.  It is illustrative to represent
a real ``angular-quantization wave'',
$$K_{i\,\ell}(m\,x) = \frac{i\,\pi}{2\,\sinh(\pi\,\ell)}\,
[I_{i\,\ell}(m\,x)-I_{-i\,\ell}(m\,x)].$$ This solution is oscillatory
for $x < \ell/m$, with a wavelength proportional to $x$, and decays
exponentially for $x > \ell/m$ (Fig.~\ref{wave}).  Actually, for $x
\ll \ell/m$, wavefunctions behave like $x^{\pm i}$, that is, like
trigonometric functions of $\ln x$.  

This type of quantization was first introduced in the context of
quantization in curved space, in particular, in Rindler space
\cite{BD}.  Rindler space is just Minkowski space and, therefore, {\em
not curved}, but in coordinates such that the time is the proper time
of a set of {\em accelerated} observers. Its remarkable feature is the
appearance of an {\em event horizon}, which implies that the ground
state (the Minkowski vacuum) is a mixed (thermal) state (the Unruh
effect).  The fact that wave-lengths vanish at $x=0$ is to be expected
from the Rindler space viewpoint, because the origin corresponds to
the horizon: the quantum states concentrate on it.  The connection
with black hole entropy and Hawking radiation is very briefly
explained in the next section.

\subsection{Black hole entropy}

The motivation to study accelerated observers was, of course, the
problem of black-hole entropy and Hawking radiation. This radiation is
perceived by static observers but not by inertial (free-falling)
observers. In fact, what a static observer close to the horizon can
see is well described by the Rindler geometry. In other words, the
large black-hole mass $M$ limit of the Schwarzschild geometry is the
Rindler geometry.  To realize this limit, it is convenient to use
the Kruskal-Szekeres coordinates $u,v$, instead of the Schwarzschild
coordinates $t,r$ ($r$ is the radial distance which together with 
time are the only relevant variables of the Schwarzschild geometry in 
any dimension) \cite{BD}.  For small values of
these coordinates (equivalently, $M \ra \infty$), the curvature
can be neglected and the geometry becomes locally the Rindler
geometry.

Once established that the geometry near the black-hole horizon is
locally the Rindler geometry of the preceding section, we can readily
transfer the form of the density matrix of a scalar field therein,
where we now ignore (trace over) the degrees of freedom inside the
horizon. Hence, we can define a von Neumann entropy associated with
this density matrix.  Furthermore, in so doing, we can appreciate that
the concept of black-hole entropy takes a new meaning: in addition to
being of {\em quantum origin}, this entropy is related to shared
properties between the interior and exterior, namely, to the
horizon.  In addition, the radial vacuum is a thermal state with
respect to the original Schwarzshild coordinates, giving rise to
Hawking radiation \cite{BD}.

\subsection{Geometric entropy}

We have seen that the half-line density matrix of a field theory has a
geometric interpretation in Rindler space. Furthermore, the entropy of
black holes can be understood as a generalization to a more
complicated (curved) geometry. Since the important feature is just the
existence of a horizon, we may wonder if further generalization is
possible.

Indeed, the notion of ``geometric entropy'' has been introduced by C.\
Callan and F.\ Wilczek \cite{CalWil}, as the entropy ``associated with
a pure state and a geometrical region by forming the pure state
density matrix, tracing over the field variables inside the region to
create an `impure' density matrix''.  They computed the Rindler space
case (like Bombelli et al \cite{Bomb}) and further proposed a
generalization to different {\em topologies}.

A different notion of geometric entropy can be deduced by purely
geometrical means from the presence of horizons, namely, as associated
with a spacetime topology that does not admit a trivial {\em
Hamiltonian foliation} \cite{HH}. This type of topology prevents
unitary evolution and produces mixed states.  In fact, it is only this
second type of entropy that leads to the famous ``one-quarter area
law'' for black holes, due to its origin in purely gravitational
concepts. On the contrary, the first notion of geometric entropy needs
an auxiliary field theory, involves UV divergences and needs
renormalization before a comparison with the gravitational notion can
be made.

\section{Conclusions}

We have seen that a density matrix renormalization group
transformation amounts to a Hilbert space reduction that essentially
preserves the information, that is, preserves the entropy.  In this
sense, it can be understood as a quantum coding operation, namely, a
quantum data compression, which is not lossless but nearly so. So a
density matrix renormalization group transformation is analogous to
standard compression of classical data, such as it is routinely used
in everyday data processing. Indeed, the singular value decomposition
is used for classical data compression when data can be arranged in
matrix form. Therefore, it is natural that it can also be applied to
quantum data compression.

The density matrix renormalization group ability to keep a constant
and small Hilbert space size while the system size grows is crucial
for its approaching a fixed point that represents the infinite size
system. This limit is necessary to study quantum phase transitions,
for example. The promising interface between quantum phase transitions
and information theory is just beginning to be studied
\cite{Vi-Lato,Scholl}.  Furthermore, in the limit of large correlation
length, the relevant dynamics, as given by the ground state and the
lowest excited states, can be described by a relativistic quantum
field theory (in which the mass of the excitations decresases with the
correlation length).  Since relativistic quantum field theories in 1+1
dimensions can be treated with powerful mathematical methods, we can
expect that they are a suitable ground to explore the connection of
quantum dyanmics with information theory.  We remark, of course, that
one can stop the renormalization group iteration at any desired point,
when some predetermined size is reached (the ``finite system method''
\cite{White}).

The density matrix renormalization group ability to keep a constant
and small Hilbert space size relies on having a distribution of
density matrix eigenvalues in which most of them are actually
negligible. In fact, their typical distribution decays exponentially.
White proposed the analogy with an ordinary statistical system with
the canonical distribution \cite{White} (which was the original
motivation of Feynman's density matrix philosophy).  We have seen that
it is more than a mere analogy: White's algorithm is equivalent to the
calculation of the density matrix in angular quantization. The
connection with the Unruh effect reveals that one can indeed associate
a particular thermodynamical picture and a temperature with angular
quantization. This picture can be generalized in terms of the concept
of geometric entropy and, in fact, connects with the notion of {\em
holography} in quantum gravity \cite{Bousso}.

Finally, regarding angular quantization and its associated
distribution of quantum states, let us remark how it helps to
understand the efficiency of White's algorithm: it is very efficient
because it employs the smallest number of boundaries allowed, namely,
just one boundary, unlike other renormalization group formulations.
For example, the block partitioning technique produces a very large
number of boundaries.
%\vskip 1cm
\bigskip

{\bf Acknowlegments:}
%\vskip 0.5cm
%{\it Acknowledgements:} 
My work is supported by 
MEC ``Ram\'on y Cajal'' Program and grant BFM2002-01014.

%I am grateful to Juan Poyatos for bringing some relevant references 
%to my attention.

\end{document}